\documentclass[aps,prb,epsfig,twocolumn]{revtex4}

\begin{document}

\bibliographystyle{prsty}

\draft


\title{Metabolic Network Modelling: Including Stochastic Effects}
\author{P. Ao}
%
%
%
\address{Department of Mechanical Engineering,
      University of Washington, Seattle, WA 98195 }
\date{ Oct. 1 (2004); revised Jan. 8 and Feb. 28, 2005 }


\begin{abstract}
We propose to model the dynamics of metabolic networks from a
systems biology point of view by four dynamical structure
elements: potential function, transverse matrix, degradation
matrix, and stochastic force. These four elements are balanced to
determine the network dynamics, which gives arise to a special
stochastic differential equation supplemented by a relationship
between the stochastic force and the degradation matrix. Important
network behaviors can be obtained from the potential function
without explicitly solving for the time-dependent solution. The
existence of such a potential function suggests a global
optimization principle, and the existence stochastic force
corresponds natural to the hierarchical structure in metabolic
networks. We provide theoretical evidences to justify our proposal
by discussing its connections to others large-scale biochemical
systems approaches, such as the network thermodynamics theory,
biochemical systems theory, metabolic control analysis, and flux
balance analysis. Experimental data displaying stochasticity are
also pointed out.
\end{abstract}


\maketitle

\section{ Introduction }

The successful completion of Human Genome Project reveals that to
understand the molecular base of diseases a huge amount of
information is needed from every level of description: from DNA,
protein, to function \cite{collins}. It becomes evident that
expertise beyond traditional biology is needed to decipher the
genomic message. In metabolic and physiological studies, there has
already been an active tradition of interaction among biologists,
chemists, physicists, engineers, and computer scientists
\cite{complexity}. Those large scale integrations call for a
rethinking on the practice of biology. The emergence of systems
biology is the response to such demand \cite{kitano}.

The systems biology study of metabolic networks may be decomposed to
be four mutually interconnected parts: \\
1). Biological experiments. It may be in response to a demand for
finding a cure for a particular disease or finding a method for
health care at the preventive level. Both labor intensive and
optimized designed experiments would be carried out. It may be an
industrial demand to increase the yield of a product, such as to
increase the production of desired biomass from methanol
\cite{lidstrom}. The better metabolic network and its related gene
regulatory network needs to be understood before a proper genetic
and metabolic engineering. Or, the goal is more on biology: we
simply wish to understand the biological principles in terms of
molecular, physical, and chemical mechanisms behind a given
biological process \cite{ideker}. To address those questions, new
tools, both experimental and computational,
as well as new theoretical framework, are needed. \\
2). Novel experimental technologies. Thus, there is an urgent
demand for better technologies, particularly in the directions of
high throughput, real time, and {\it in vivo} at cellular level.
Those technologies should provide us new interrogative power to
understand biological mechanism, with new and more data.  \\
3). Biocomputation and bioinformatics.  The new and large scale
data sets call for their proper annotations and presentations, the
task of biocomputation and bioinformatics.
There have been tremendous efforts recently in this direction. \\
4). Conceptual and mathematical frameworks.  Hence, all those
new developments call for more powerful conceptual
and mathematical frameworks to facilitate a better
communications among experts of theoretical and
experimental biologists and to help achieve the goals.  \\
The purpose of the present article is to address the question of
mathematical framework from a biophysics and biochemistry point of
view, motivated by our study in systems biology, particularly in
the theoretical study of a gene regulatory network \cite{zhu1} and
in the experimental study of a metabolic network \cite{lidstrom}.

In addition to large scale biological data sets, more time
sequence data have been obtaining and {\it in vivo} data plays
increasingly important roles. They often show that data are
stochastic in nature \cite{division1,division2,oshea}, not due to
our inability to obtain accurate data. They carry molecular and
cellular biological information. We further notice that the
biological processes are physical and chemical. This implies that
we must work with constraints, such mass conservation on biomass,
the energy conservation, as well as other thermodynamic
constraints. Hence there exist several requirements for the
modelling: We demand it to provide insights and understandings on
bio-structure, to perform exploring simulations, to interpret and
evaluate of measured data, to make prediction and help design
further experiments, and to optimize the whole process to achieve
the goal.  For those ambitious objectives, we believe a proper
mathematical framework may be based on the stochastic differential
equations. This is a middle level description: It can be linked to
both the master equation type description with explicit discrete
nature of chemical processes and the Fokker-Planck equation type
with continuous variables in both time and state space. One
advantage of present description is that a very specific form can
be spelt out: the four dynamical elements formulation of
stochastic differential equations. They can be directly connected
to experimental data. The role played by stochascitity is
emphasized in this description.

In section II a discussion of the dynamical components will be
given. In section III the specific form will be given and its
connection to more classical form of stochastic differential
equation in chemistry is obtained. In section IV a further
discussion of dynamical elements in sight of quantitative
equations will be given. Applications are discussed in section V.
In section VI the connection of present framework to several
successful approaches in metabolic network modelling, the network
thermodynamics theory, biochemical systems theory, metabolic
control analysis, and flux balance analysis. We conclude in
section VII.

\section{Four Dynamical Components in Metabolic Networks}

Because biological processes are based on physics and chemistry,
we would naturally assume that they are continuous in time.
Further, because we are focused on the metabolic networks, we
would like to assume that the most important ingredient in the
mathematical description would be a set of coupled chemical
reaction rate equations, which may be presented by a set of
stochastic differential equations. In fact, this has been the
working assumption underlying all the mathematical metabolic
network modelling \cite{san}. This middle level description can
link to other descriptions of either the complete discrete master
equation type or of the complete continuous Fokker-Planck equation
type. The important question is whether or not a more useful
mathematical structure exists.

To address the dynamical structure question of a given network
describing the macromolecular synthesis in metabolic pathways, we
may intuitively classify its dynamical components into four basic
elements according to their roles in the providing or spending of
the resource in a metabolic network: the driving force, the
transverse force, the degradation, and the noise. There has
already been lots of previous attempts in this direction in the
content of non-equilibrium thermodynamics \cite{nicolis}. The
driving force is the active part in such macromolecular synthesis,
which would include the external force and the internal structure
determined by the metabolic network architecture. It provides the
necessary resource, such as the energy, the mass flow, or other
needed chemical ingredients, in order to maintain the desired
function of the network. It is closely related to the robustness
of the network, acting as a backbone or skeleton of network
dynamics. It is the gradient of the potential function of the
network. The transverse force describes the process of relocation
of the resource from one part of the network to another, and is
responsible for phenomena such as the delay response and the
oscillation. The effects of both driving and transverse forces
alone are reversible: no change of the potential function of the
network. The dissipation describes how the resource is consumed
during the network dynamical process, including the consumption of
metabolites, dissipation of energy, and degradation of
macromolecules. It is an irreversible process, and cannot be
avoided for a complex metabolic network. Owing to dissipation, the
network cannot maintain its designed function if external and
internal resources are finished. Finally, for any complex network,
there always exist factors that cannot be known for certain or
whose detailed dynamical behavior is not our primary interest.
Those factors form the fourth element: the noise or the stochastic
force. Both dissipation and noise give the network the ability to
adapt to the optimal state. We will return for further discussion
of those four elements in Section IV.

Interestingly, a complex network has only these four dynamical
elements within present mathematical framework. Their interplaying
gives rise to rich behaviors which we will explore elsewhere. In
the present article, we will present the underlying quantitative
mathematical framework for the dynamical structure in terms of two
generic dynamical principles: The equivalence principle that all
the four elements must be balanced to describe the network
dynamics; The stochasticity-dissipation relation that a constraint
exists on the stochastic force. Such a constraint implies that the
stochastic force must be the integral part of the network
dynamics, a view most evident in biology \cite{luria}. The
explicit identification of the four dynamical elements of a
network leads to a better understanding of the network behavior:
The final stationary distribution can be directly read out from
the driving force; the time scales in the network becomes
explicit; and a long sought optimization principle for a network
is immediately suggested.

\section{Stochastic Differential
    Equations to Describe Metabolic Network Dynamics}

The questions naturally arise that how do we describe the network
dynamics quantitatively and how do we identify the four elements?
To be specific, let us consider an $n$ component network. The $n$
components may be the relevant the numbers of metabolites in a
metabolic network or the relevant proteins in a gene regulatory
network pathway \cite{zhu1} or other quantities specifying the
network The value of $j^{th}$ component is denoted by $q_j$. The
$n$ dimensional vector ${\bf q}^{\tau} = (q_1, q_2, ... , q_n)$ is
the state variable of the network. Here the superscript $\tau$
denotes the transpose. Let $f_j({\bf q})$ be the deterministic
nonlinear force on the $j^{th}$ component, which includes both the
effects from other components and itself, and $\zeta_j( {\bf q},
t)$ the random force. For simplicity we will assume that $f_j$ is
a smooth function explicitly independent of time. The network
dynamics may be generally modelled by the set of stochastic
differential equations, the stochastic chemical rate equations
\cite{san,nicolis}:
\begin{equation}
  \dot{q}_{j} = f_j( {\bf q} ) + \zeta_j( {\bf q} , t),
\end{equation}
with $j = 1, 2, ..., n $. Here $\dot{q}_{j} = d q_{j}/d t$.
Without loss of generality the noise will
be assumed to be Gaussian and white with the variance,
\begin{equation}
    \langle \zeta_i ( {\bf q} , t) \zeta_j ({\bf q}, t')  \rangle
  = 2 D_{ij}( {\bf q} ) \delta (t-t') ,
\end{equation}
and zero mean, $   \langle \zeta_j   \rangle = 0$. Here
$\delta(t)$ is the Dirac delta function, and $  \langle ...
\rangle $ indicates the average with respect to the dynamics of
the stochastic force. By the physics and chemistry convention the
semi-positive definite symmetric matrix $D=\{D_{ij}\}$ with
$i,j=1,2, ..., n$ is the diffusion matrix. Eq.(1) is the standard
stochastic differential equation, or the standard stochastic
differential equation in physics and chemistry. If an average over
the stochastic force ${\bf \zeta}$, a Wiener noise, is performed,
Eq.(1) is reduced to the deterministic equation in dynamical
systems, $   \langle \dot{\bf q}   \rangle =   \langle {\bf
f}({\bf q})  \rangle = {\bf f}(  \langle {\bf q}   \rangle)$.

With Eq.(1) and (2) as our starting point to identify the
dynamical elements, three immediate problems need to be addressed.
First, the stochastic force ${\bf \zeta}$ appears to be simply
introduced by hand. It is not obvious how it can be related to the
deterministic force ${\bf f}$. Second and more serious, apart from
the apparent {\it ad hoc} stochastic force, in Eq.(1) there are no
sign of other three dynamical elements: the driving force, the
transverse force, and the dissipation. Third and worse, both the
divergence and the skew matrix of the force ${\bf f}$ are in
general finite:
\begin{equation}
 \partial \cdot {\bf f} \neq 0, \;
 \partial \times {\bf f} \neq 0 \;.
\end{equation}
Here the divergence is explicitly $\partial \cdot {\bf f} =
\sum_{j=1}^{n} \partial f_j /\partial q_j = tr(F)$, and the skew
matrix is twice the anti-symmetric part of the force matrix $F$: $
( \partial \times {\bf f} )_{ij} = F_{ji} - F_{ij}$ with $ F_{ij}
= \partial f_i /\partial q_j \ , \; i,j  =1,2, ..., n$. The
finiteness of the divergence leads to that the phase space volume
is not conserved. Dissipation is hence implied. The finiteness of
the skew matrix has been characterized as the hallmark of the
network in a state far away from thermal equilibrium
\cite{nicolis}.

We propose, and have demonstrated elsewhere \cite{ao}, that there
exists a decomposition such that Eq.(1) can be transformed into
the following form:
\begin{equation}
  [ S({\bf q}) + T ({\bf q}) ] \dot{\bf q}
     = - \partial u ({\bf q}) + {\bf \xi}({\bf q}, t) \; ,
\end{equation}
with the semi-positive definite symmetric matrix $S$, the
anti-symmetric matrix $T$, the single-valued scalar function $u$,
and the stochastic force ${\bf \xi}$. Here $\partial$ is the
gradient operator in the state variable space. It is
straightforward to verify that the symmetric matrix term is
`degradation':
\[
  \dot{ \bf q}^{\tau} S({\bf q}) \dot{\bf q} \geq 0 \; ;
\]
 and the anti-symmetric part does no `work':
\[
  \dot{\bf q}^{\tau} T ({\bf q}) \dot{\bf q} = 0 \; ,
\]
therefore non-decaying. Hence, it is natural to identify that the
degradation is represented by the symmetric matrix $S$, the
friction matrix, and the transverse force by the anti-symmetric
matrix $T$, the transverse matrix. The driving force is clearly
represented by the gradient of the scalar function $u$, which
acquires now the meaning of potential energy and will be named the
network function. There are indeed four dynamical elements for the
network dynamics. Eq.(4) states that they must be balanced in
order to produce the network dynamics.

Eq.(4) is our key for the identification of dynamical structure.
It is the form of stochastic differentiation equation which we
believe would be suitable to directly analyze metabolic networks.
Nevertheless, without further constraint, the four dynamical
elements, if exists, are not unique. This may be illustrated by a
simple counting: There are four apparent independent quantities in
Eq.(4), while there are only two in Eq.(1). For this reason the
decomposition from Eq.(1) to (4) will be called the singular
decomposition. In order to have a unique identification, we choose
to impose the constraint on the stochastic force and the symmetric
matrix in Eq.(4) in the following manner:
\begin{equation}
    \langle {\bf \xi}({\bf q},t) {\bf \xi}^{\tau} ({\bf q},t')   \rangle
    = 2 S ({\bf q}) \delta(t-t') \; ,
\end{equation}
and $   \langle {\bf \xi}({\bf q})  \rangle = 0$. This constraint
is consistent with the Gaussian and white noise assumption for
${\bf \zeta}$ in Eq.(1), and will be called the
stochasticity-dissipation relation. The constrained singular
decomposition will be called the gauged singular decomposition.
Eq. (4) and (5) will be called the normal stochastic differential
equation, or the normal stochastic differential equation.

Two comments are in order. First, it is evident that any matrix
can be decomposed into a symmetric and an antisymmetric matrices.
What is unique in Eq.(4) is that the symmetric matrix in Eq.(4)
has to be semi-positive definite, as guaranteed by Eq.(5).
Therefore, the matrix $[S+T]$ at left hand side of Eq.(4) is not
an arbitrary matrix. It has a built-in structure which makes the
network tend to the minimum of potential function $u({\bf q})$,
that is, the tendency for optimization. Since the potential
function $u({\bf q})$ plays the same role as energy function in
physics and chemistry, the typical optimization procedure, such as
the simulation annealing, may be used here to find the global
minimum of the network potential function. This leads to the
second comment that without the stochastic effect in Eq.(1), no
unique potential function in Eq.(4) can be determined. This
implies that no global optimization can be found in the absence of
stochastic effect.

The connection from Eq.(1), the typical form of chemical rate
equation, to Eq.(4), the form of the present proposal, may be
expressed in following equations \cite{ao}:
\begin{equation}
  \left\{ \begin{array}{lll}
  u & = & - \int_C d{\bf q}' \cdot
          [ G^{-1}({\bf q}') {\bf f}({\bf q}') ] \\
  S & = & [G^{-1}({\bf q}) + (G^{\tau} )^{-1}({\bf q}) ] /2 \\
  T & = & [G^{-1}({\bf q}) - (G^{\tau} )^{-1}({\bf q}) ] /2
  \end{array}
     \right. \; .
\end{equation}
Here the

Here is the auxiliary matrix function $  G({\bf q}) = [S({\bf q})
+ T({\bf q}) ]^{-1} $  is the solution of following equations:
\begin{equation}
  \partial \times [ G^{-1} {\bf f}({\bf q}) ] = 0 \; ,
\end{equation}
and
\begin{equation}
  G + G^{\tau} = 2 D \; .
\end{equation}

With the network function $u({\bf q})$ similar to a potential
energy, the stationary distribution function $\rho_0({\bf q})$ for
the state variable is expected to be a type of Boltzmann-Gibbs
distribution:
\begin{equation}
   \rho_0({\bf q}) = \frac{1}{Z} \; \exp\{ - u({\bf q}) \} \; ,
\end{equation}
with the partition function $Z = \int d^n {\bf q} \; \exp\{ -
u({\bf q}) \} $. This is one of most useful results from Eq.(4):
It allows a direct comparison to stochastic experimental data at
steady states.

\section{Further Clarification}

With above clear identification of four dynamical elements in a
complex network from Eq.(4) and the demonstration of its
equivalence to the tradition formulation of Eq.(1), a further
discussion of them is given in this section.

$\xi({\bf q},t)$. We start with the stochastic force $\xi({\bf
q},t)$, the noise. It is intuitively evident that such a force
always exists. In a more mathematical description, this force can
come either from the environmental influence on the network, or
from approximations such that the continuous representation of a
discrete process. Here a Gaussian and white noise is assumed. In
reality, more complicated noises, non-Gaussian and colored, can
exist. The presentation formulation in the form of Eq.(4) should
provide a good starting point. For example, Eq.(4) is already in
the form in dissipative dynamics\cite{leggett} where colored
noises have been readily considered. The emphasis on noise in the
presentation formulation also suggests that metabolic fluxes may
not be good dynamical variables. The better variables may be the
numbers of proteins or other macromolecules inside the cell which
carry out the metabolic task.

$S({\bf q}) \dot{\bf q}$. The friction matrix $S({\bf q})$ and the
frictional force $S \dot{\bf q}$ are self-evident, too. The
existence of the friction shows that the network has the tendency
to approach to a steady state. The friction can be the real
friction in mechanics, or is a representation of the degradation
of proteins or other materials in biology. The present dynamical
structure theory requires that the friction is always associated
with the noise according to the stochasticity-dissipation
relation, Eq.(5). The friction and noise are the two opposite
sides of stochastic dynamics: the ability to adaptation with
friction and the ability to optimization with noise.

$T({\bf q}) \dot{\bf q}$. The antisymmetric matrix $T({\bf q})$
and the transverse force $ T \dot{\bf q} $ are less obvious. In
physical sciences, the antisymmetric matrix is closely related to
quantities in physics and chemistry such as the mass, the magnetic
field, and the rotation. It is a general statement on the
conjugate relations among state variables. The well-known examples
of corresponding transverse forces are Lorentz and gyroscopic
forces. In a complex network, the transverse force represents the
network ability to relocate the resource from one part of the
network to another. In this sense it is similar to the conversion
of energy between the kinetic energy and potential energy in
mechanics, with the aid of a finite mass. The transverse force is
responsible for oscillatory behaviors in networks.

$u({\bf q})$. The driving force $ - \partial u({\bf q}) $ is the
gradient of the scalar function $u({\bf q})$, the network
potential, with respect to the network state variable ${\bf q}$.
The network potential is the most important quantity, determining
the robustness of network dynamics. Knowing this network potential
is equivalent to knowing the landscape in which the network
evolves: Its minima determine local equilibrium states, and its
absolute minimum is the ultimate steady state of the network.
Therefore important dynamical properties of the network can be
read from $u$ without explicit solving for the time dependent
solutions.

We should point out that the potential $u({\bf q})$ so defined is
an emerging property of the metabolic network. It has no direct
thermodynamic meaning as those in physics and chemistry.
Specifically, it is not the usual free energy in thermodynamics.
Since the potential $u({\bf q})$ describes what the network would
eventually like to be under all thermodynamic and other
constraints, it is a quantity at a higher level description than
those of free energies to describe the network reaction rates.
Similar dynamical structures as those in Eq.(4) and (5) also exist
in other branches of biology \cite{ao2004b}.

\section{Illustrations}

In this section we give two illustrations, one mathematical and
one biological. We first discuss a useful approximation scheme to
find Eq.(4) from Eq.(1). In this way the connection between
present formulation, Eq.(4), and the classical formation in
biochemistry, Eq.(1), becomes more clear. Then we brief summarize
an successful example of the application the four dynamical
element analysis in an outstanding stability puzzle of a gene
regulatory network.

\subsection{First order gradient expansion}

We give an explicit demonstration of how to obtain Eq.(4) from
Eq.(1) by a so-called gradient expansion to the first order
derivative of force ${\bf f}$ with respect to the state variables.
To be specific, we only consider a two dimensional problem. Here
$q_1$ and $q_2$ represent numbers of two enzymes in a cell. The
force in Eq.(1) consists of two effects, the production rate and
the degradation rate:
\begin{equation}
    f_i({\bf q}) = f_{ip}({\bf q}) - f_{id}({\bf q}) \; \; i=1,2 \; ,
\end{equation}
with the subscripts $p$ and $d$ stand for the production and
degradation respectively. Under the diffusion approximation, the
stochastic force is \cite{vankampen,mcquarrie}
\begin{equation}
 \zeta_i({\bf q},t) = \sqrt{ f_{ip}({\bf q}) } \zeta_{ip}(t) +
                      \sqrt{ f_{id}({\bf q}) } \zeta_{id}(t)
                         \; \; i=1,2 \; ,
\end{equation}
with $\zeta_{ip}(t), \zeta_{id}(t)$ are unity random variables and
possible correlation among them. Therefore the diffusion matrix
$D$ can be readily obtained, which is what needed below. We remark
that the equation similar to above has been used in the study of
bio-networks \cite{wolynes,zhu1,oshea}.

The construction of Eq.(4) from Eq.(1) will be given to the lowest
order in the gradient expansion. The usefulness of this
approximated construction can be illustrated for following three
reasons. First, in many practical applications, this approximation
is a good approximation \cite{zhu1}. In fact, it is exact in the
linear case. Second, the Eq.(7) becomes algebraic, instead of a
partial differential equation. The complete solution of such
algebraic equation can be found \cite{kat}. Third, several salient
features of the gauged decomposition becomes apparent without
undue mathematical complications. An important quantity is the
force matrix $F$. According to the definition following Eq.(3),
\begin{equation}
    F_{11} = \partial_1 f_1 \, ,    F_{12} = \partial_2 f_1 \, ,
    F_{21} = \partial_1 f_2 \, ,    F_{22} = \partial_2 f_2  \; .
\end{equation}
Eq.(8) will not change under the gradient approximation. In the
lowest order gradient approximation, Eq.(7) becomes simple. We
collect them here:
\begin{equation}
  \left\{ \begin{array}{lll}
   G F^{\tau} - F G^{\tau} & = & 0  \\
   G          + G^{\tau}   & = & D
  \end{array}
     \right. \; .
\end{equation}
In two dimensions the matrix manipulation is particularly
straightforward. The antisymmetric part of the auxiliary matrix $G
= D + Q$ from Eq.(13) can be found to be
\begin{equation}
  Q = { (F D - D F^{\tau}) }/ { tr(F) } \; .
\end{equation}
Using the relation
\[
   \left( \begin{array}{ll}
                 M_{11} & M_{12} \\
                 M_{21} & M_{22}
    \end{array} \right) ^{-1} = \frac{1}{ \det(M) }
    \left( \begin{array}{ll}
                 M_{22} & - M_{12} \\
               - M_{21} & M_{11}
    \end{array} \right)
\]
The friction matrix $S$ and the antisymmetric matrix $T$ can be
found according to Eq.(6):
\begin{equation}
  \left\{ \begin{array}{lll}
    u & = & - \int_C d{\bf q}' \cdot
          [ G^{-1}({\bf q}') {\bf f}({\bf q}') ] \\
    S & = &
         \left( \begin{array}{ll}
                 D_{22} & - D_{12} \\
               - D_{12} & D_{11}
         \end{array} \right)  /\det(D)  \\
    T & = & {-Q}/{\det(G)}
  \end{array}
     \right. \; .
\end{equation}
In two dimensions, $\det(G) = \det(D) + \det(Q) = D_{22}D_{11} -
 D_{12}^2 + Q_{12}^2 $ and is obviously non-negative.

\subsection{Solving the stability puzzle of a genetic switch}

We have used the scheme here to study the outstanding puzzle of
stability and efficiency of phage lambda genetic switch in a
quantitative manner \cite{zhu1,zhu2}. The key was to find as
accurate as possible the potential function of the gene
maintenance regulatory network. The dynamical equation for such a
network can be written down according to chemical rate equations.
Information at three levels, the DNA, protein, and function, needs
to be integrated to provide a quantitative model. Based on above
approach we are able to obtain the potential function and achieve
a quantitative agreement with biological data. We are able to make
quantitative predictions such as the amount of cooperative free
energy. We show mathematically that a robust and efficient genetic
switch can be obtained: potential function is a measure of
robustness and stochastic force is responsible for the efficiency.

We should remark that in reality, both the intrinsic and the
extrinsic noise coexist. They are equally important and measurable
\cite{oshea}. The stochasticity-dissipation relation treats them
on the equal footing to determine the gauged singular
decomposition. We have made used of this fact in the modelling of
noise in our effort to solve the stability puzzle of the genetic
switch \cite{zhu1,zhu2}.

\section{ Connections to Known Metabolic Network Modelling}

In section III  we demonstrated that the proposed four dynamical
component formulation of stochastic differential equations is
equivalent to the usual chemical rate equations. The proposed
method offers a direct connection to observed data because of the
transparent meaning of the potential and its role in the
optimization. In the following we briefly discuss its connections
to other successful modelling methodology in metabolic network
study.

1). Flux Balance Analysis (FBA). There are two ingredients in FBA
\cite{fba}: The first one is the mass conservation, built into the
formulation through the stoichiometric consideration. The second
one is a linearizing near a steady state state. The first feature
is very useful in that it eliminates redundant dynamical
variables, which would be Eq.(1) more tractable. The second
feature may be viewed as a special case of Eq.(1), by setting the
stochastic force to be zero, and by linearizing the force ${\bf
f}$ around its stable fixed point with the given stoichiometric
constraints. The justification for the stable fixed point is
natural because of the existence of heomostasis: The steady state
of the biological process should be stable under the given
constraints \cite{fba}.

Thought it is in general a rather weak constraint, FBA has been
most easy to implement in practical applications, because many
matured mathematically tools, such as linear programming, can be
employed. It is currently the working horse in modelling and
engineering of metabolic networks, augmented by additional
considerations \cite{wk}.

2). Metabolic Control Analysis (MCA). To go beyond the simple
linear case and to study how the network in steady state responds
to changes in fluxes and its components have been the primary goal
of MCA \cite{fell}. Properties of the architectural structure of
the metabolic network can be revealed by MCA.  Undoubtly, MCA may
be viewed as the study of the structure built into the force ${\bf
f}$ in Eq.(1), the situation without dynamics and stochastic
force.

3). Biochemical Systems Theory (BST). A more general and well
grounded approach is BST \cite{voit}: The force ${\bf f}$ is
assumed to polynomial to assist mathematical analysis. Time
dependent is included into the formulation but the stochastic
force is typically neglected. Methodologies in cybernetics or
control theory are employed in BST. This approach is clearly from
an engineer's point of view, providing a great insight for the
metabolic network engineering.

4). Stoichiometric Network Thermodynamic theory (SNT).  With the
concern that there is a tendency to ignore the thermodynamic
constraints in metabolic network modelling, the recently developed
SNT attempts to remedy such shortcomings with explicitly
incorporation of such constraints \cite{qian}. Indeed, because
energy and entropy play such dominant roles in metabolic network
dynamics, this is a very desirable progress. Novel results have
been obtained along this approach \cite{qian2,qian3}. Again, the
starting point of SNT is an equation with the same form as of
Eq.(1), which shows that our framework is compatible with SNT,
too. It remains to find the connection between the potential
function in Eq.(4) and those of thermodynamic quantities employed
in SNT.

We should point out that in the present article it is impossible
to give an adequate survey of vast literature on the modelling of
metabolic networks. Fortunately, several excellent books already
exist \cite{san,fba,wk,fell,voit,tyson,cortassa}, where ample
discussions on FNA, MCA, BST and others can be found.  The need to
incorporate stochastic effects into modelling has already been
demonstrated by stochasticity in biological experimental data
\cite{zhu1,division1,division2,oshea,mcadams}.

Another feature should also be pointed here. From the present
stochastic modelling, Eq.(4) or (1), there are obviously two time
scales: the very short one characterizing the stochastic force
$\xi({\bf q},t)$ or $\zeta({\bf q},t)$ and the time scale on which
the smooth functions of potential function $u(\bf q)$, degradation
(friction) matrix $S({\bf q})$ and the transverse matrix $T({\bf
q})$ having well defined meaning. This corresponds nicely to the
hierarchical structure abundant in metabolic pathway analysis
\cite{reich}.

\section{Conclusion}

In this paper we have proposed a general systems approach to model
metabolic network dynamics. It is based on a special form of
stochastic differential equations.  We have demonstrated that it
is equivalent to the classical chemical rate equations approach
with noise. The connections to various existing modelling
methodologies are pointed out: In each of their valid description
regions our method is compatible. Hence it may provide a unifying
mathematical framework with the particular useful potential
function. Its usefulness has already been demonstrated in the
study of a gene regulatory network. It remains to be demonstrated
that additional biological insights in metabolic network study can
be obtained. This is the task been undertaking.

{\ }

{\bf Acknowledgements.} First of all I would to like to thank M.E.
Lidstrom for introducing me to the wonderful field of metabolism
and for sharing with me her biological insights. Helpful
discussions with G. J. Crowther, S.M. Stolyar, and S. van Dien to
clarify biological issues and with G. Kosaly and H. Qian to
clarify mathematical issues were gratefully acknowledged. This
work was supported in part by a USA NIH grant under the number
HG002894.


\begin{thebibliography}{99}

\bibitem{collins}
  F.S. Collins, E.D. Green, A.E. Guttmacher, and M.S. Guyer, A vision for the future of
     genomics research. Nature 422 (2003) 835-847.
\bibitem{complexity}
  Complexity in Biological Information Processing, Novartis Foundation Symposium
     239, edited by G. Bock and J. Goode. John Wiley and sons, New York, 2001.
\bibitem{kitano}
  H. Kitano, Systems biology: A brief overview.  Science 295 (2002) 1662-1664.
\bibitem{lidstrom}
   L. Chistoserdova, S.W. Chen, A. Lapdius, and M.E. Lidstrom,  Methylotrophy in
      {\it Methylobacterium extorquens} AM1 from genomic point of view.
       J. Bacterio. 185 (2003) 2980-2987.
\bibitem{ideker}
  T. Ideker, T. Galitski, and L. Hood, A new approach to decoding
      life: systems biology. Annu. Rev. Genomics Hum. Genet. 2 (2001)
       343-372;
\bibitem{zhu1}
   X.-M. Zhu, L. Yin, L. Hood, and P. Ao, Calculating biological behaviors of epigenetic
        states in the phage lambda life cycle. Funct. Integr. Genomics 4 (2004) 188-195.
\bibitem{division1}
  S. Umehara, Y. Wakamoto, I. Inoue, and K. Yasuda, On-chip single-cell
      microcult6ivation assay for monitoring environmental effects on isolated cells.
      Biochem. Biophy. Res. Commun. 305 (2003) 534-540.
\bibitem{division2}
   A. Elfwing, Y. LeMarc, J. Baranyi, and A. Ballagi, Observing growth and division
        of large numbers of individual bacteria by image analysis.
        Appl. Environ. Microbiol. 70 (2004) 675-678.
\bibitem{oshea}
   J.M. Raser and E.K. O'shea, Control of stochasticity in eukaryotic gene expression.
      Science 304 (2004) 1811-1814.
\bibitem{san}
  G. N. Stephanopoulos, A.A. Aristidou, and J. Nielsen, Metabolic Engineering:
     principles and methodologies. Academic Press, San Diego, 1998.
\bibitem{nicolis}
   G. Nicolis and I. Prigogine, Self-Organization in Nonequilibrium Systems:
      from dissipative structure to order through fluctuations.
      John Wiley and sons, New York, 1977.
\bibitem{luria}
   S.E. Luria and M. Delbruck, Mutations of bacteria from virus sensitivity
      to virus resistance. Genetics 28 (1943) 491-511.
\bibitem{ao}
   P. Ao, Potential in stochastic differential equations: novel construction.
     J. Phys. A37 (2004) L25-L30.
\bibitem{leggett}
  A.J. Leggett, Quantum tunneling of a macroscopic variable, pp1-36,
     in Quantum Tunnelling in Condensed Media, edited by
    Yu. Kagan and A.J. Leggett. North-Holland, Amsterdam, 1992.

\bibitem{ao2004b}
  P. Ao, Mathematical structure of evolutionary theory.
    arXicv eprint: q-bio/0403020.
%
\bibitem{vankampen}
   N.G. van Kampen, Stochastic Processes in Physics and Chemistry.
   Elsevier, Amsterdam, 1992.
\bibitem{mcquarrie}
   D.A. McQuarrie, Stochastic approach to chemical kinetics.
      J. Appl. Prob. 4 (1967) 413-478.

\bibitem{wolynes}
  M. Sagai and P.G. Wolynes, Stochastic gene expression as a many-body problem.
      Proc. Nat'l. Acad Sci USA 100 (2003) 2374-2379.
\bibitem{kat}
  C. Kwon, P. Ao, and D.J. Thouless, Structure of stochastic dynamics near fixed points.
       (submitted to PNAS.)
\bibitem{zhu2}
   X.-M. Zhu, L. Yin, L. Hood, and P. Ao,  Robustness, stability and efficiency of
      phage lambda genetic switch: dynamical structure analysis. J. Bioinform. Comput.
     Biol. 2 (2004) 785-817.

\bibitem{fba}
  J.S. Edwards et al, in  Metabolic Engineering, edited by S.Y. Lee
      and E.T. Papoutsakis. Marcel Dekker, New York, 1999.
\bibitem{wk}
  Metabolic Engineeing in the Post Genomic Era, edited by B.N.
      Kholodenko and H.V. Westerhoff. Horizon Bioscience, Norfolk,
      2004.
\bibitem{fell}
  D.A. Fell, Understanding the Control of Metabolism. Portland
      Press, London, 1996.
\bibitem{voit}
  E.O. Voit,  Computational Analysis of Biochemical Systems: a
     practical guide for biochemists and molecular biologists.
     Cambridge University Press, Cambridge, 2000.
\bibitem{qian}
  H. Qian and D.A. Bear, Thermodynamics of stoichiometric
     biochemical networks far from equilibrium.
     Biophysical Chemistry (2005) (in press).
\bibitem{qian2}
  D.A. Beard, H. Qian, and S.D. Liang, Energy balance for analysis
     of complex metabolic networks. Biophys. J. 83 (2002) 79-86.
\bibitem{qian3}
  H. Qian, D.A. Beard, and S.D. Liang, Stoichiometric network theory
      for nonequilibrium biochemical systems. Eur. J. Biochem. 270 (2003)
      415-421.
\bibitem{tyson}
  C. P. Fall, E.S. Marland, J.M. Wagner, and J.J. Tyson, Computational
      Cell Biology. Springer-Verlag, Berlin, 2002.
\bibitem{cortassa}
  S. Cortassa, M.A. Aon, A.A. Iglesias, and D. Lloyd, An
      Introduction to Metabolic and Cellular Engineering.
      World Scientific, Singapore, 2002.
\bibitem{mcadams}
  A. Arkin, J. Ross, and H.H. McAdams, Stochastic kinetic analysis
      of developmental pathway bifurcation in phage lambda-infected
      Escherichia coli cells. Genetics 149 (1998) 1633-1648.
\bibitem{reich}
   J.G. Reich and E.E. Selkov, Energy Metabolism of the Cell: a
   theoretical treatise. Academic Press, New York, 1981.


\end{thebibliography}
\end{document}